\documentclass[twocolumn,preprintnumbers,amsmath,amssymb,prb,floatfix,nofootinbib,superscriptaddress]{revtex4}

\usepackage{graphicx,tabularx}
\usepackage[version=3]{mhchem} 
\usepackage{amssymb}
\usepackage{dcolumn}
\usepackage{color}
\usepackage[mathcal]{euscript}
\usepackage{color}
\usepackage[title,titletoc,toc]{appendix}
\definecolor{gray0}{gray}{0.0}
\definecolor{gray64}{gray}{0.25}
\definecolor{gray128}{gray}{0.5}
\definecolor{gray192}{gray}{0.75}
\definecolor{gray255}{gray}{1.0}
\definecolor{darkgreen}{rgb}{0,0.6,0}

\def\Ce3{Ce$^{\mathrm{III}}$}
\def\La3{La$^{\mathrm{III}}$}
\def\Ce4{Ce$^{\mathrm{IV}}$}
\def\A3{A$^{\mathrm{III}}$}
\def\B4{B$^{\mathrm{IV}}$}

\def\CO2{CeO$_2$}
\def\C2O3{Ce$_2$O$_3$}
\def\Ef{E$_f$}

\begin{document}

\title{Tetravalent doping of \ce{CeO2}: The impact of valence electron character on group IV dopant influence.}
\author{D. E. P. Vanpoucke}
\affiliation{Department of Inorganic and Physical Chemistry, Ghent University, Krijgslaan $281$ - S$3$, $9000$ Gent, Belgium}
\affiliation{Center for Molecular Modeling, Ghent University, Technologiepark $903$, $9052$ Zwijnaarde, Belgium}
\email[corresponding author:]{Danny.Vanpoucke@ugent.be}
\author{S. Cottenier}
\affiliation{Center for Molecular Modeling, Ghent University, Technologiepark $903$, $9052$ Zwijnaarde, Belgium}
\affiliation{Department of Materials Science and Engineering, Ghent University, Technologiepark $903$, $9052$ Zwijnaarde, Belgium}
\author{V. Van Speybroeck}
\affiliation{Center for Molecular Modeling, Ghent University, Technologiepark $903$, $9052$ Zwijnaarde, Belgium}
\author{I. Van Driessche}
\affiliation{Department of Inorganic and Physical Chemistry, Ghent University, Krijgslaan $281$ - S$3$, $9000$ Gent, Belgium}
\author{P. Bultinck}
\affiliation{Department of Inorganic and Physical Chemistry, Ghent University, Krijgslaan $281$ - S$3$, $9000$ Gent, Belgium}

\date{\today}
\begin{abstract}
Fluorite \ce{CeO2} doped with group IV elements is studied within the DFT and DFT+U framework. Concentration dependent formation energies are calculated for  \ce{Ce_{1-$x$}Z_{$x$}O2} (Z= C, Si, Ge, Sn, Pb, Ti, Zr, Hf) with $0\leq x \leq 0.25$ and a roughly decreasing trend with ionic radius is observed. The influence of the valence and near valence electronic configuration is discussed, indicating the importance of filled $d$ and $f$ shells near the Fermi level for all properties investigated. A clearly different behavior of group IVa and IVb dopants is observed: the former are more suitable for surface modifications, the latter are more suitable for bulk modifications.\\
\indent For the entire set of group IV dopants, there exists an inverse relation between the change, due to doping, of the bulk modulus and the thermal expansion coefficients. Hirshfeld-I atomic charges show that charge transfer effects due to doping are limited to the nearest neighbor oxygen atoms.
\end{abstract}

\maketitle
\section{Introduction}
\indent Over the last two decades, interest in ceria based materials has grown steadily. The facile \ce{Ce^{4+} \rightleftharpoons} \ce{Ce^{3+}} redox conversion in combination with the fact that \ce{CeO2} retains its fluorite crystal structure even with significant degrees of reduction makes ceria based materials ideal for controlling the oxygen atmosphere in three-way catalysts.\cite{TrovarelliA:CatalRev1996, KasparJ:CatalToday1999, FuQ:2003Science, SheYusheng:IJHE2009}
Furthermore, the introduction of dopants like Zr has been shown to increase the oxygen storage capacity, increasing the range of practical applicability.\cite{DeLeitenburgC:JChemSocChemCommun1995}
The remarkable redox properties of \ce{CeO2} make ceria based materials also of interest in several other industrially important applications, such as: catalytic support, thermal barrier coatings, ionic conductors and fuel cells.\cite{KundakovicLj:JCat1998, ManzoliMaela:CT2008, LiB:IJHE2010, VanpouckeDannyEP:2011PhysRevB_LCO} More recently, CeO$_2$ and ceria based materials have been used as thin film buffer layers in \ce{YBa2Cu3O_{7-$\delta$}} coated superconductors.\cite{ParanthamanM:1997PhysC, OhSanghyun:PhysC1998, PennemanG:2004EuroCeram, TakahashiY:PhysC2004, KnothKerstin:PhysC2005, VandeVeldeNigel:EurJInorChem2010, VyshnaviN:2012JMaterChem}\\
\indent For all these applications there is a constant search for ways to improve the properties of ceria based materials even further. One of the most common ways for such improvements is through the introduction of dopants, with regimes ranging from light doping ($<1$\%) to mixed oxides containing $50$\% dopants or more.\cite{TrovarelliA:CatalRev1996,MogensenM:SolStateI2000} Different applications also require different properties to be optimized. Unfortunately, optimization towards one application may worsen the performance of the system for another. As an example, for ionic conductors one wishes to maximize the ionic conductivity, while for a buffer layer one often wishes to minimize this quantity. Such opposing and varying requirements make a more general theoretical study of the influence of dopants on the properties of \ce{CeO2} of general interest. This is the particular aim of the present paper.\\
\indent Experimentally, lanthanide elements are most often studied as dopants, in addition to standard catalyst $d$-block metals (Au, Pd, Cu, ...).\cite{KudoT:JES1975, McBrideJR:JApplPhys1994, TrovarelliA:CatalRev1996, KundakovicLj:ApplCatalA1998, RossignolS:JMaterChem1999, MogensenM:SolStateI2000, BeraP:ChemMater2002, YamamuraH:JCSJ2003, FuQ:2003Science, WangX:JPhysChemB2005, FaggDP:JSolStateChem2006, TiwariA:ApplPhysLett2006, NakamuraA:PureApplChem2007, SongYQ:JApplPhys2007, VodungboB:ApplyPhysLett2007, deBiasi:JAlloysCompd2008, ManzoliMaela:CT2008, LiB:IJHE2010, SinghalRK:JPhysDApplPhys2011} Since all these elements can act as aliovalent dopants, charge compensating oxygen vacancies are introduced, contributing to the changes in the \ce{CeO2} properties. To have a clear understanding of the influence dopants have on the properties of \ce{CeO2}, it is important to separate the contributions of these two structural changes; \textit{i.e.} to distinguish between the consequences of doping and the subsequently introduced charge compensating vacancies. Theoretical calculations are ideally suited to separate the different contributions. Due to the tetravalent nature of Ce in \ce{CeO2},  group IV elements are the perfect starting point for such investigations.\\
\indent Only few studies addressed group IV dopants in ceria based materials, although for example silica is a widely used catalyst support.\cite{ReddyBM:JPhysChemB2005,ReddyBM:CatalSurvAsia2005} This limited experimental work originates from the difficulty of forming solid solutions with \ce{CeO2} in a controlled way. Improved redox properties compared to pure and Zr doped \ce{CeO2} have been presented for mixed \ce{CeO2}-\ce{SiO2} and \ce{CeO2}-\ce{SnO2} materials.\cite{RocchiniE:JCatal2000, RocchiniE:JCatal2002, LinR:ApplCatal2003, ReddyBM:JPhysChemB2005} Andersson \textit{et al.} studied the redox properties of tetravalent dopants at low dopant concentrations of $3$\% and found a correlation between the reducibility and the dopant's ionic radius.\cite{AnderssonDA:2007bPhysRevB, AnderssonDA:2007ApplPhysLett} Tang \textit{et al.} investigated the redox properties of tetravalent dopants (Mn, Pr, Sn and Zr) and linked the lower oxygen vacancy  formation energy to structural distortions and electronic modifications.\cite{TangYuanhao:PhysRevB2010} Yashima studied tetragonal crystal phases of \ce{Ce_{1-x}Zr_{x}O2} and found a tetragonal phase to appear for Zr concentrations $\geq 37.5$\%, and a cubic phase with displaced O atoms at lower concentrations.\cite{YashimaMasatomo:JPhysChemC2009} We previously found Vegard law behavior for group IV doped \ce{CeO2} and proposed the Shannon crystal radius instead of the ionic radius as parameter for the lattice expansion.\cite{VanpouckeDannyEP:2012aApplSurfSci}\\
\indent In the present paper, we investigate the modification of the mechanical properties of \ce{CeO2} due to doping with group IV elements using density functional theory (DFT).
We show there is a difference in energetics for doping between group IVa and IVb elements, and link this to the different valence electronic structure. (Sec.~\ref{CeO2p4:ssc_ResultsEf}--\ref{CeO2p4:ssc_ResultsDOS}). An inverse relation  between the bulk modulus and the thermal expansion coefficients is presented (Sec.~\ref{CeO2p4:ssc_ResultsBMTEC}) and the change in the charge distribution around the dopants discussed (Sec.~\ref{CeO2p4:ssc_ResultsHI}). Technical details of the computational methods used are provided in appendix~\ref{CeO2p4:sc_theormeth}.

\begin{table}[!tbp]
\caption{Crystal structure of the reference bulk and oxide materials used.}\label{table:Suppl1_GeomBulkMetal}
\begin{ruledtabular}
\begin{tabular}{l|cccccc}
 & structure & Pearson Symbol & space group \\
 &           &                &             \\
\hline\\[-2mm]
C  & diamond & cF$8$ &  Fd$\overline{3}$m  \\
Si & diamond & cF$8$ &  Fd$\overline{3}$m \\
Ge & diamond & cF$8$ &  Fd$\overline{3}$m  \\
Sn & diamond & cF$8$ &  Fd$\overline{3}$m  \\
Pb & face centered cubic & cF$4$ &  Fm$\overline{3}$m  \\
\hline\\[-2mm]
Ti & hexagonal close packed & hP$2$ & P$63$/mmc \\
Zr & hexagonal close packed & hP$2$ & P$63$/mmc \\
Hf & hexagonal close packed & hP$2$ & P$63$/mmc \\
\hline\\[-2mm]
CO$_2$  & dry ice         & cP$12$ &  Pa$\overline{3}$  \\
SiO$_2$ & $\alpha$-quartz & cF$8$  &  P$3_121$ \\
GeO$_2$ & rutile (argutite)& tP$6$  &  P$4_2$/mnm  \\
SnO$_2$ & rutile (cassiterite)& tP$6$  &  P$4_2$/mnm  \\
PbO$_2$ & rutile (plattnerite)& tP$6$  &  P$4_2$/mnm  \\
\hline\\[-2mm]
TiO$_2$ & Anatase         & tI$12$ & I$4_1$/amd \\
ZrO$_2$ & Baddeleyite     & mP$12$ & P$12_1$/c$1$ \\
HfO$_2$ & monoclinic(Baddeleyite)& mP$12$ & P$12_1$/c$1$ \\
\end{tabular}
\end{ruledtabular}
\end{table}

\begin{table}[!btp]
\caption[Defect formation energy for group IV dopants]{Defect formation energy for group IV dopants at different concentrations.}\label{table:Metal4subst_energies}
\begin{ruledtabular}
\begin{tabular}{l|cccc|c}
 & \multicolumn{5}{c}{\Ef (eV)} \\
 &  25\% & 12.5\% & 3.704\% & 3.125\% & 3.125\%\\
\hline
 & \multicolumn{4}{c|}{LDA} & LDA+U$^{a}$ \\
\hline
CeO$_2$ & \multicolumn{4}{c|}{-11.484$^{b}$} & --\\
\hline
C & $20.435$ & $20.654$ & $20.747$ & $20.748$ & $21.088$  \\
Si & $8.288$ & $8.078$ & $8.050$ & $8.052$ & $8.358$  \\
Ge & $9.404$ & $9.134$ & $9.110$ & $9.105$ & $9.341$  \\
Sn & $7.061$ & $6.866$ & $6.851$ & $6.851$ & $7.013$  \\
Pb & $8.963$ & $8.757$ & $8.742$ & $8.745$ & $8.866$  \\
\hline
Ti & $3.168$ & $3.254$ & $3.248$ & $3.233$ & $3.563$  \\
Zr & $0.530$ & $0.555$ & $0.548$ & $0.543$ & $0.796$  \\
Hf & $0.105$ & $0.139$ & $0.131$ & $0.128$ & $0.392$  \\
\hline
 & \multicolumn{4}{c|}{PBE} & PBE+U$^{a}$\\
\hline
CeO$_2$ & \multicolumn{4}{c|}{-10.418$^{b}$} & -- \\
\hline
C & $19.687$ & $19.889$ & $19.954$ & $19.967$ & $20.143$  \\
Si & $8.737$ & $8.547$ & $8.530$ & $8.562$ & $8.764$   \\
Ge & $9.495$ & $9.257$ & $9.242$ & $9.254$ & $9.386$   \\
Sn & $7.129$ & $6.967$ & $6.962$ & $6.960$ & $7.031$   \\
Pb & $8.960$ & $8.788$ & $8.777$ & $8.780$ & $8.812$   \\
\hline
Ti & $3.454$ & $3.526$ & $3.524$ & $3.532$ & $3.740$   \\
Zr & $0.851$ & $0.884$ & $0.885$ & $0.883$ & $1.033$   \\
Hf & $0.484$ & $0.517$ & $0.517$ & $0.517$ & $0.666$   \\
\end{tabular}
\end{ruledtabular}
\begin{flushleft}
$^{a}$ For the DFT+U calculations a Hubbard U$=5$ eV was chosen both for the LDA and PBE calculations.\\
$^{b}$ Instead of the formation energy the heat of formation is given.\\
\end{flushleft}
\end{table}

\section{Results and Discussion}
\subsection{Defect formation energies}\label{CeO2p4:ssc_ResultsEf}
\indent The stability of the different group IV doped fluorite \ce{CeO2} systems is investigated through the comparison of the defect formation energies \Ef, which are calculated as:
\begin{equation}\label{eq:matal4subst_Eform}
E_{f} = E_{Ce_{1-x}Z_xO_2} - E_{CeO_2} + N_{df}(E_{Ce}-E_{Z}),
\end{equation}
with $N_{df}$ the number of substituent atoms in the system, $E_{Ce_{1-x}Z_xO_2}$ the total energy of the relaxed doped system, $E_{CeO_2}$ the total energy of a \ce{CeO2} fluorite supercell of equal size, $E_{Ce}$ and $E_{Z}$ the energy per atom of bulk $\alpha$-Ce and the bulk phase of the substituent element $Z$. Larger positive values of $E_{f}$ indicate that more energy is required for doping. Pure Ce metal has several different phases depending on the external pressure and temperature. In our case we chose to use the $\alpha$-phase, since the calculations are performed at zero temperature and pressure. The used crystal structure of the dopant element bulk phase is presented in appendix \ref{CeO2p4:sc_theormeth}, Table~\ref{table:Suppl1_GeomBulkMetal}. Defect formation energies can also be calculated with regard to dopant-oxides, and are presented in appendix \ref{CeO2p4:Appendix_sc_Ef}. The obtained results remain qualitatively the same as those presented below.\\
\indent Table~\ref{table:Metal4subst_energies} shows that $E_{f}$ varies only slightly with dopant concentration. For cells with a dopant concentration of $3.125$\% $E_{f}$ is also calculated within the DFT+U framework, using the LDA/PBE optimized geometry and a Hubbard U$=5$ eV for the Ce $f$ electrons. As can be seen in Table~\ref{table:Metal4subst_energies}, the formation energies are slightly larger in the DFT+U framework, although they do not change the relative stability of the systems compared to the pure LDA/PBE results.\\
\indent For different substituent elements the formation energies show a spread over quite a large range, with roughly $20$ eV/atom required to imbed a C atom, down to merely $0.1$ eV/atom required to imbed a Hf atom. This shows that none of the group IV elements increase the stability of \ce{CeO2}. A  direct comparison with experiment is not straightforward, since \textit{ab initio} calculations do not account for effects of pressure and temperature. In addition, during the experimental preparation of a compound there are often several steps (thermal and/or mechanical treatment, \ldots) involved which provide additional energy to the system that enable the formation of metastable doped systems. For example, in experiments one observes that oxygen vacancies form spontaneously in \ce{CeO2} cubic fluorite,\cite{TrovarelliA:CatalRev1996} although \textit{ab initio} simulations show them always to require energy. In an attempt to compare with experiment, we use our calculated oxygen vacancy formation energy as a reference. Substitutions that require the same \textit{ab initio} computed energy as this reference, or less, are considered more likely under experimental conditions, while those requiring more, less likely. For the LDA and PBE functionals we calculated the O vacancy formation energy to be $4.035$ and $3.097$ eV per vacancy, respectively, at a vacancy concentration of $1.5625$\%, in excellent agreement with values found in literature.\cite{AnderssonDA:2007aPhysRevB,GandugliaPirovanoMVeronica:SurfSciReports2007} Comparing the $4.035$ eV found for pure LDA to the $3.61$ eV for LDA+U presented by
Andersson \textit{et al.},\cite{AnderssonDA:2007aPhysRevB} shows that the introduction of a
Hubbard U term does not qualitatively alter the stability difference observed for
the group IVa and IVb dopants. Note, however, that the pure LDA value is closer to the
experimental value of $4.65$--$5.00$ eV for bulk reduction.\cite{PanhansMA:SolStateIon1993} For higher O vacancy concentrations higher formation energies are found, up to $5.006$ and $4.145$ eV for LDA and PBE respectively, at $12.5$\%, which is within the range of experimentally derived values.\cite{PanhansMA:SolStateIon1993}\\
\begin{figure}
  \includegraphics[width=8cm,keepaspectratio=true]{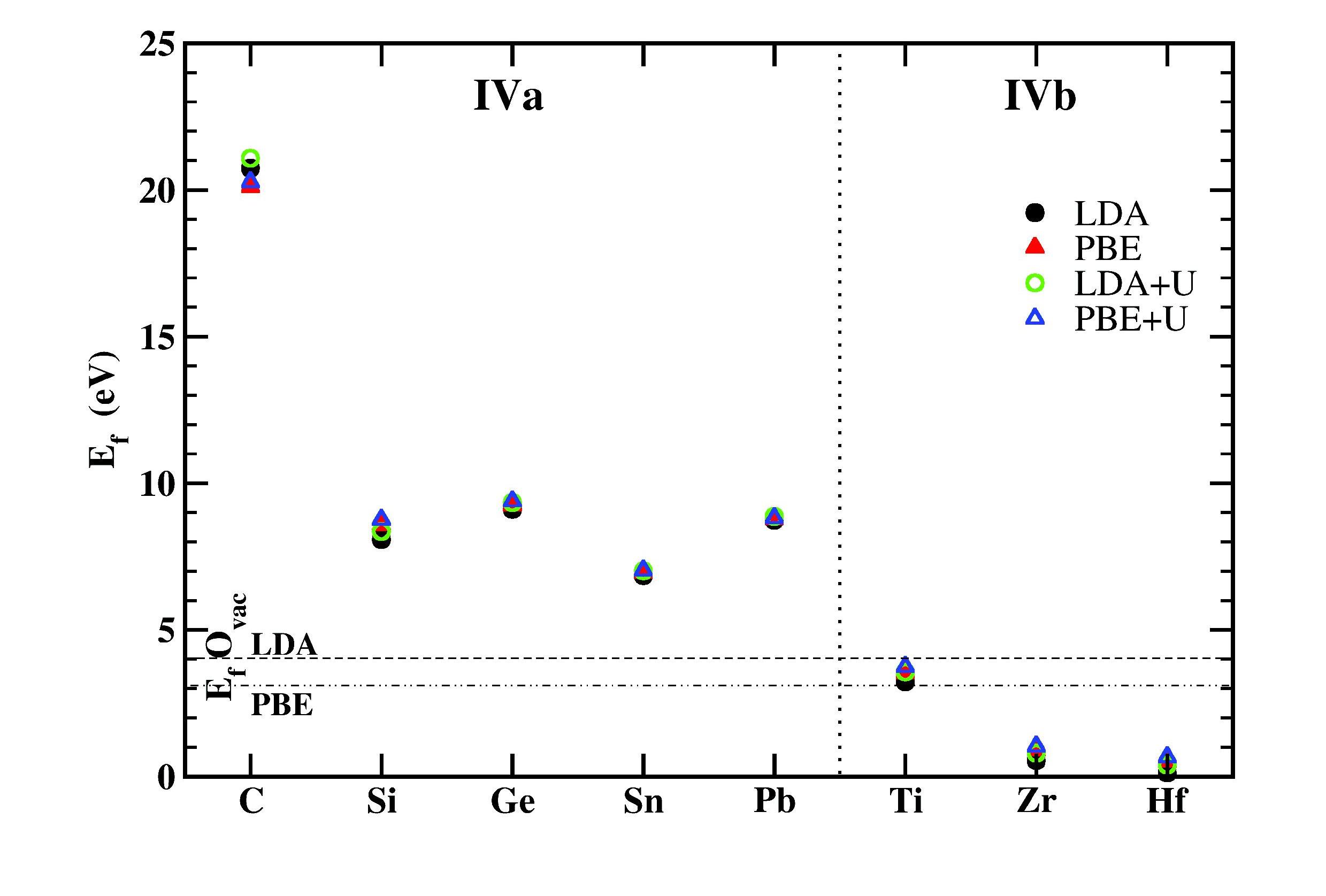}\\
  \caption{(Color online) The formation energy E$_f$ for group IV doped \ce{CeO2}, at a dopant concentration of $3.125$\%. LDA (LDA+U) values are indicated as black discs (green circles), while PBE (PBE+U) values are indicated as red solid (blue empty) triangles. The oxygen vacancy formation energy (E$_{\mathrm{f}}$ O$_{\mathrm{vac}}$) for $1.56$\% of oxygen vacancies is indicated as a dashed and a dashed double dotted line for LDA and PBE, respectively.}\label{fig:EfVSEl}
\end{figure}
\indent Based on these reference energies the dopants in Table~\ref{table:Metal4subst_energies} nicely split in the ``more likely bulk dopants'' (group IVb) and the ``less likely bulk dopants'' (group IVa). In this context, group IVb dopants are likely to remain (homogeneously) dispersed in the bulk of the \ce{CeO2} grains and crystals, while the group IVa dopants would likely segregate to the surface regions of the \ce{CeO2} grains, or the interface regions with other materials, or cluster to form small domains of the dopant element inside \ce{CeO2} crystals and grains. This makes the latter of interest for applications where tuning of surface effects is important. The former, however, are expected to remain distributed in the \ce{CeO2} bulk, making them well suited for applications where \ce{CeO2} bulk parameters need to be modified, or in oxide mixing experiments. These results also agree with the observation that Zr-doped \ce{CeO2} is widely used and easily produced in experiments, while it is much harder to form Si and Sn doped \ce{CeO2} in a controlled way.\cite{ReddyBM:JPhysChemB2005, RocchiniE:JCatal2000, RocchiniE:JCatal2002, LinR:ApplCatal2003}\\
\indent For the group IVb elements, the formation energy decreases with increasing atomic size, showing a better size-match with Ce results in reduced system strain.\cite{VanpouckeDannyEP:2012aApplSurfSci} The formation energies for the group IVa doped systems show a similar globally decreasing trend. The additional oscillation is interesting, and is retained even with the introduction of the Hubbard U correction. The slight increase in the formation energy for Ge and Pb coincides with the introduction of a `new' filled shell, the $d$ shell in case of Ge, and the $f$ shell in case of Pb. This shows that not only the outer valence electrons play a role in the stability of the system but also the presence of filled shells near the Fermi level.\\

\begin{figure}
\begin{center}
    \includegraphics[width=8cm,keepaspectratio=true]{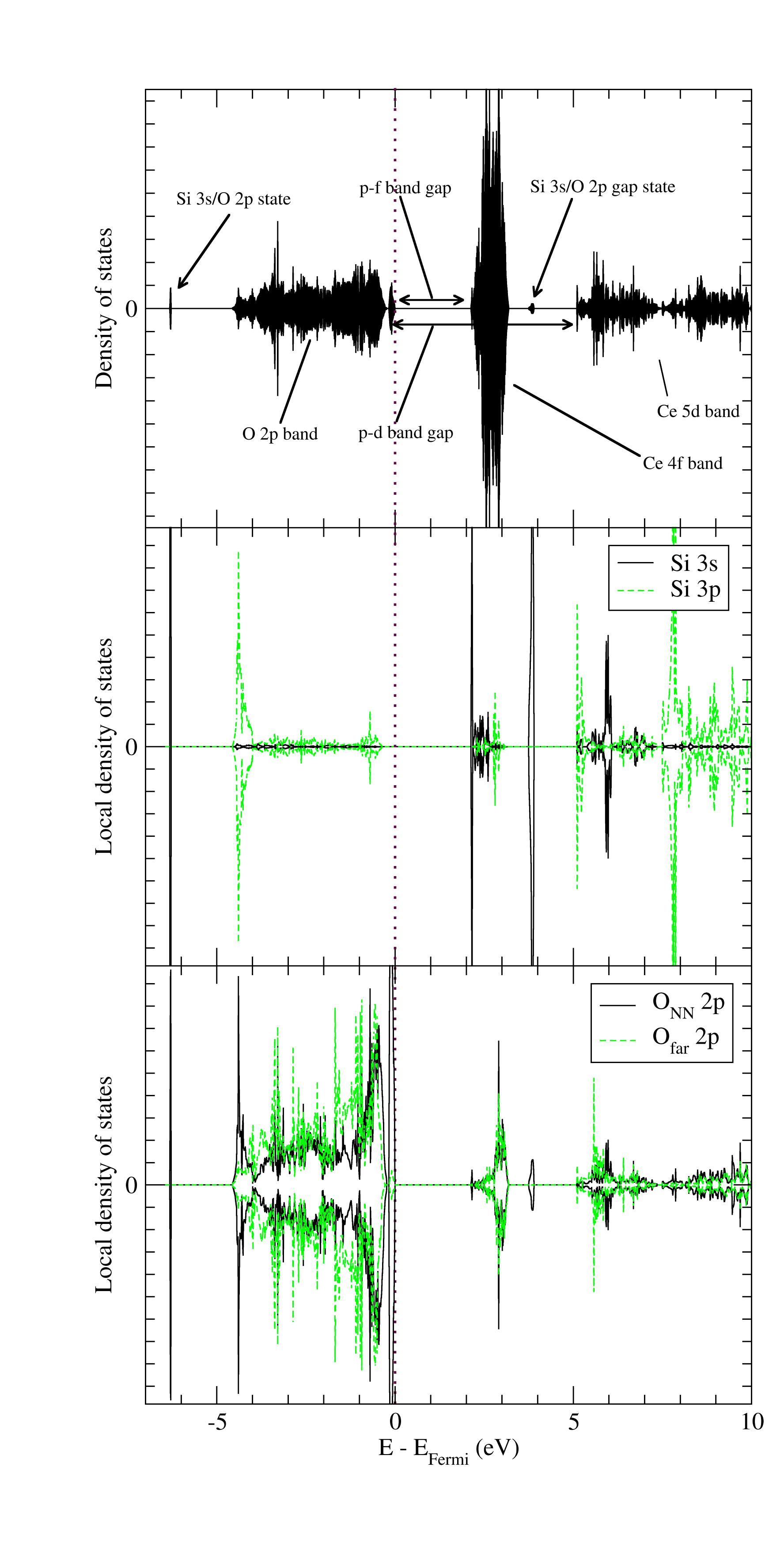}\\
\end{center}
  \caption[The total DOS for \ce{Ce_{0.96875}Si_{0.03125}O2}]{(Color online) The total DOS for \ce{Ce_{0.96875}Si_{0.03125}O2}, indicating important features (top). The $3s$ and $3p$ Si local density of states, showing the origin of the gap state and atomic state below the O $2p$ band (middle). The $2p$ O local density of states for the $8$ nearest neighbor (NN) O atoms surrounding the Si dopant, and a O atom far away from the dopant (bottom). The upper and lower panels present the spin up and spin down channels, respectively. The calculations were performed within the DFT+U framework.}\label{fig:DOS_Si}
\end{figure}
\begin{figure}
\begin{center}
  \includegraphics[width=8cm,keepaspectratio=true]{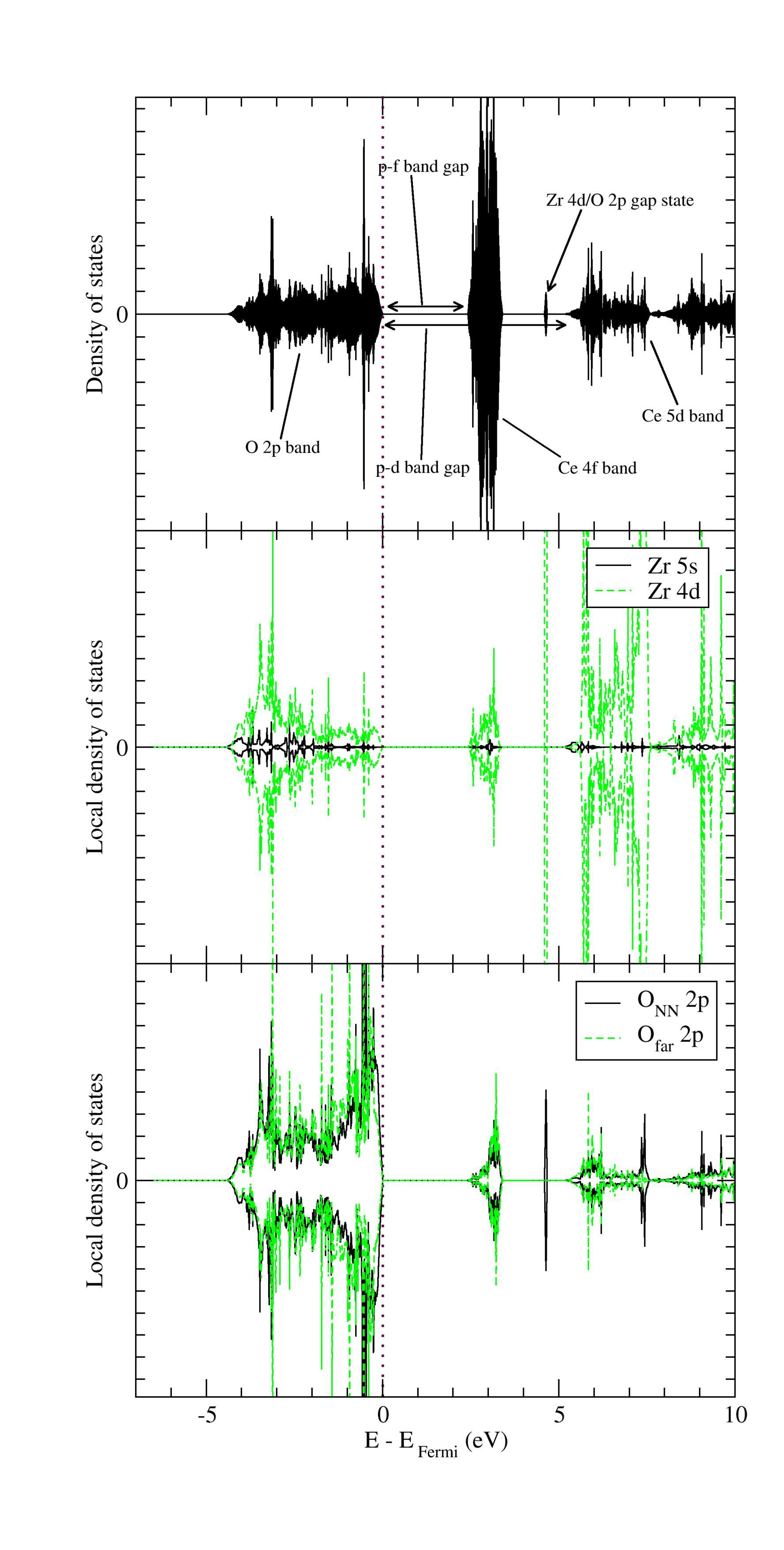}\\
\end{center}
  \caption[The total DOS for \ce{Ce_{0.96875}Zr_{0.03125}O2}]{(Color online) The total DOS for \ce{Ce_{0.96875}Zr_{0.03125}O2}, indicating important features (top). The $5s$ and $4d$ Zr local density of states, showing the origin of the gap state and atomic state below the O $2p$ band (middle). The $2p$ O local density of states for the $8$ nearest neighbor (NN) O atoms surrounding the Zr dopant, and a O atom far away from the dopant (bottom). The upper and lower panels present the spin up and spin down channels, respectively. The calculations were performed within the DFT+U framework.}\label{fig:DOS_Zr}
\end{figure}
\begin{table}[!btp]
\caption[Band gap sizes in group IV doped \ce{CeO2} ]{Band gaps in group IV doped \ce{CeO2}.$^{a}$}\label{table:Metal4subst_DOSdata}
\begin{ruledtabular}
\begin{tabular}{l|cccccc}
 & \multicolumn{3}{c}{LDA+U}  & \multicolumn{3}{c}{PBE+U}\\
 & $2p$--$4f$ & $2p$--$5d$ & gs & $2p$--$4f$ & $2p$--$5d$ & gs \\
\hline
\ce{CeO2} & $2.45$ & $5.35$ & -- & $2.37$ & $5.35$ & -- \\
\hline
C  & $2.26$ & $5.18$ & $-0.17$ & $2.12$ & $5.14$ & $-0.21$ \\
Si & $2.14$ & $5.11$ & $3.87$  & $2.06$ & $5.10$ & $3.62$ \\
Ge & $2.24$ & $5.16$ & $1.80$  & $2.18$ & $5.19$ & $1.67$ \\
Sn & $2.30$ & $5.25$ & $3.88$  & $2.22$ & $5.26$ & $3.65$ \\
Pb & $2.38$ & $5.29$ & $1.79$  & $2.29$ & $5.29$ & $1.61$ \\
\hline
Ti & $2.36$ & $5.02$ & $4.18$  & $2.27$ & $5.05$ & $3.82$ \\
Zr & $2.42$ & $5.19$ & $4.63$  & $2.33$ & $5.19$ & $4.44$ \\
Hf & $2.41$ & $5.48$ & $5.37$  & $2.32$ & $5.43$ & $5.25$
\end{tabular}
\end{ruledtabular}
\begin{flushleft}
$^{a}$ Size of the \ce{Ce_{1-$x$}Z_{$x$}O2} (with $x=0.03125$) band gaps between the O $2p$ and Ce $4f$ bands and between the O $2p$ and Ce $5d$ bands (\textit{cf.}~Figs.~\ref{fig:DOS_Si} and \ref{fig:DOS_Zr}). The position of the gap state (gs) with regard to the valence band edge is given for the group IVa elements ($s$ state), and the group IVb elements ($d$ state). All values are given in eV.\\
\end{flushleft}
\end{table}
\subsection{Density of states}\label{CeO2p4:ssc_ResultsDOS}
\indent We investigated the influence, at low concentration ($x=3.125$\%), of group IV dopants on the density of states (DOS) of \ce{Ce_{1-$x$}Z_{$x$}O2}, within the DFT+U framework. Due to the low concentration, the general shape of the DOS is comparable to the DOS of pure \ce{CeO2}; there is a conduction band of unoccupied Ce $4f$ states in the band gap between the O $2p$ valence band and Ce $5d$ conduction band (\textit{e.g.} Fig.~\ref{fig:DOS_Si} and \ref{fig:DOS_Zr} ). Table~\ref{table:Metal4subst_DOSdata} shows that the band gap between the unoccupied Ce $4f$ states and the O $2p$ valence band is always smaller than for pure \ce{CeO2}. Excluding C, we find for both group IVa and IVb dopants an increase in the band gap size with the atomic number (within each group). The same behavior is observed for the O $2p$ -- Ce $5d$ band gap. Comparison of the LDA+U and PBE+U results in Table~\ref{table:Metal4subst_DOSdata}, shows that the trends are independent of the functional used. In case of the group IVa elements, there is an atomic band roughly $2$ eV below the O $2p$ band (\textit{e.g.} Fig.~\ref{fig:DOS_Si}). From the local DOS (LDOS) it is clear that this localized state is a combination of the dopant $s$ state and the O $2p$ state of the O ions surrounding the dopant. This is shown for Si doping in Fig.~\ref{fig:DOS_Si}, and agrees well with the work of Andersson \textit{et al.}\cite{AnderssonDA:2007bPhysRevB} where a symmetry broken configuration was studied. In contrast to that work we also observe such a localized combined state in the $2p$--$5d$ band gap (\textit{cf.}~Fig.~\ref{fig:DOS_Si}). Closer investigation reveals that both bands each integrate to two electrons, indicating that this is a pair of bonding and anti-bonding states. To investigate this discrepancy the LDOS for a symmetry broken configuration
(\ce{Ce_{0.96875}Si_{0.03125}O2}) was calculated. With the exception of the \ce{Ce} $4f$
gap state, due to the presence of an oxygen vacancy in the structure of
Andersson \emph{et al.},\cite{AnderssonDA:2007bPhysRevB}
we obtained qualitatively the same LDOS as was reported by these authors. This indicates that
our observed $s$--$p$ gap state originates from the different chemical environment for the
Si dopant. However, it remains unclear to us if this state simply vanished upon symmetry
breaking, or merely merged with the Ce $5d$ band. The work of Tang \textit{et al.} on Sn doped \ce{CeO2} also shows the appearance of such atomic states below the O $2p$ band and in the band gap.\cite{TangYuanhao:PhysRevB2010} Furthermore, Tang \textit{et al.} show that the gap state lies under the Fermi level when an oxygen vacancy next to the dopant is created, showing the excess electrons to transfer to the group IVa dopant. Table~\ref{table:Metal4subst_DOSdata} shows the position of these gap states with regard to the valence band edge. This shows that only for C doping this state is occupied and is located at the edge of the valence band (hence the negative value). For the other IVa elements the band is either located above or below the unoccupied Ce $4f$ band. The relation between these gap states and the atomic state below the O $2p$ valence band becomes even clearer due to the strong correlation of their position, placing the atomic states for the Ge and Pb doped systems roughly $0.7$ eV below those of the Si and Sn doped systems.\\
\indent In the LDOS of the heavier group IVa elements also filled $d$ (Ge, Sn) and $f$ (Pb) states near the Fermi level are observed, in line with the behavior expected from the calculated formation energies (\textit{cf.} Table~\ref{table:Metal4subst_energies}), gap state positions (\textit{cf.} Table~\ref{table:Metal4subst_DOSdata}) and mechanical properties (\textit{cf.} Table~\ref{table:Metal4subst_BM_TEC}).\\
\indent For the group IVb elements no atomic state below the O $2p$ valence band is present, only a gap state above the unoccupied Ce $4f$ band is observed (\textit{e.g.}~Fig.~\ref{fig:DOS_Zr}). This state moves toward the Ce $5d$ band, with increasing ionic size of the dopant (\textit{cf.}~Table~\ref{table:Metal4subst_DOSdata}). For the Ti and Zr dopants the band integrates to $6$ and $4$ electrons, respectively. In case of the Hf dopant this band shows a small overlap with the Ce $5d$ band, and the number of electrons is estimated to be in the range of that found in Ti and Zr doped ceria. Just as for the IVa elements, this localized state is a combination with O $2p$ states of the O ions surrounding the dopant, as is shown for Zr in Fig.~\ref{fig:DOS_Zr}.\\
\indent Additionally, the DOS and LDOS of systems with a dopant concentration of $25$\% were investigated within the pure DFT framework. We found qualitatively similar results, showing the presence of $d$ and $f$ states just below the Fermi-level for group IVa dopants with filled $d$ and $f$ shells.\\
\indent For the group IVa dopants (except C), the presence of $s$-type conduction states results in a serious reduction of the $2p$--$4f$ band gap, compared to the low concentration systems. Furthermore, the position of the maximum of these $s$ bands appears correlated with the defect formation energies.\\
\indent In case of the group IVb dopants, on the other hand, the $2p$--$4f$ band gap does not change with the dopant concentration. Similar as for the low concentration systems, the unoccupied Ce $4f$ band contains a significant contribution of the group IVb dopant $d$-state (\textit{cf.}~Fig.~\ref{fig:DOS_Zr}).\\
\indent As a result, the clearly better stability of the group IVb doped systems, may be attributed to the better resemblance of the dopant LDOS to the LDOS of Ce. In addition, the reduction of the $2p$--$4f$ band gap width in the group IVa doped systems, and its correlation to the defect formation energy suggests that either the $2p$--$4f$ band gap is important for the stability of the system, or that the presence of $s$ states near the Fermi-level is detrimental for its stability.
\begin{table}[!btp]
\caption[Calculated bulk moduli and thermal expansion coefficients for group IV doped \ce{CeO2}]{Calculated bulk moduli and thermal expansion coefficients for group IV doped \ce{CeO2}.$^a$}\label{table:Metal4subst_BM_TEC}
\begin{ruledtabular}
\begin{tabular}{l|cccc}
& \multicolumn{2}{c}{$B_0$ (Mbar)} &  \multicolumn{2}{c}{$\alpha$ ($10^{-6}$K$^{-1}$)} \\
& LDA & PBE & LDA & PBE \\
\hline
\ce{CeO2} & $2.017$ & $1.715$ & $11.218$ & $12.955$ \\
\hline
C  & $1.528$ & $1.235$ & $16.287$ & $20.379$ \\
Si & $2.057$ & $1.738$ & $12.335$ & $14.994$ \\
Ge & $1.909$ & $1.573$ & $14.390$ & $18.454$ \\
Sn & $2.004$ & $1.692$ & $12.157$ & $14.357$ \\
Pb & $1.845$ & $1.516$ & $14.006$ & $17.715$ \\
\hline
Ti & $2.145$ & $1.825$ & $11.195$ & $12.971$ \\
Zr & $2.153$ & $1.878$ & $10.960$ & $12.212$ \\
Hf & $2.194$ & $1.881$ & $10.652$ & $12.146$
\end{tabular}
\end{ruledtabular}
\begin{flushleft}
$^{a}$ The bulk modulus (BM) $B_0$ for \ce{CeO2} is calculated for a dopant concentration of $25$\%. The linear TEC $\alpha$ is calculated at the same dopant concentration and a temperature of $500$ K. The full TEC curves are shown in Fig.~\ref{fig:Subst_ThermEC_X4}.\\
\end{flushleft}
\end{table}

\begin{figure}
\begin{center}
  \includegraphics[width=8cm,keepaspectratio=true]{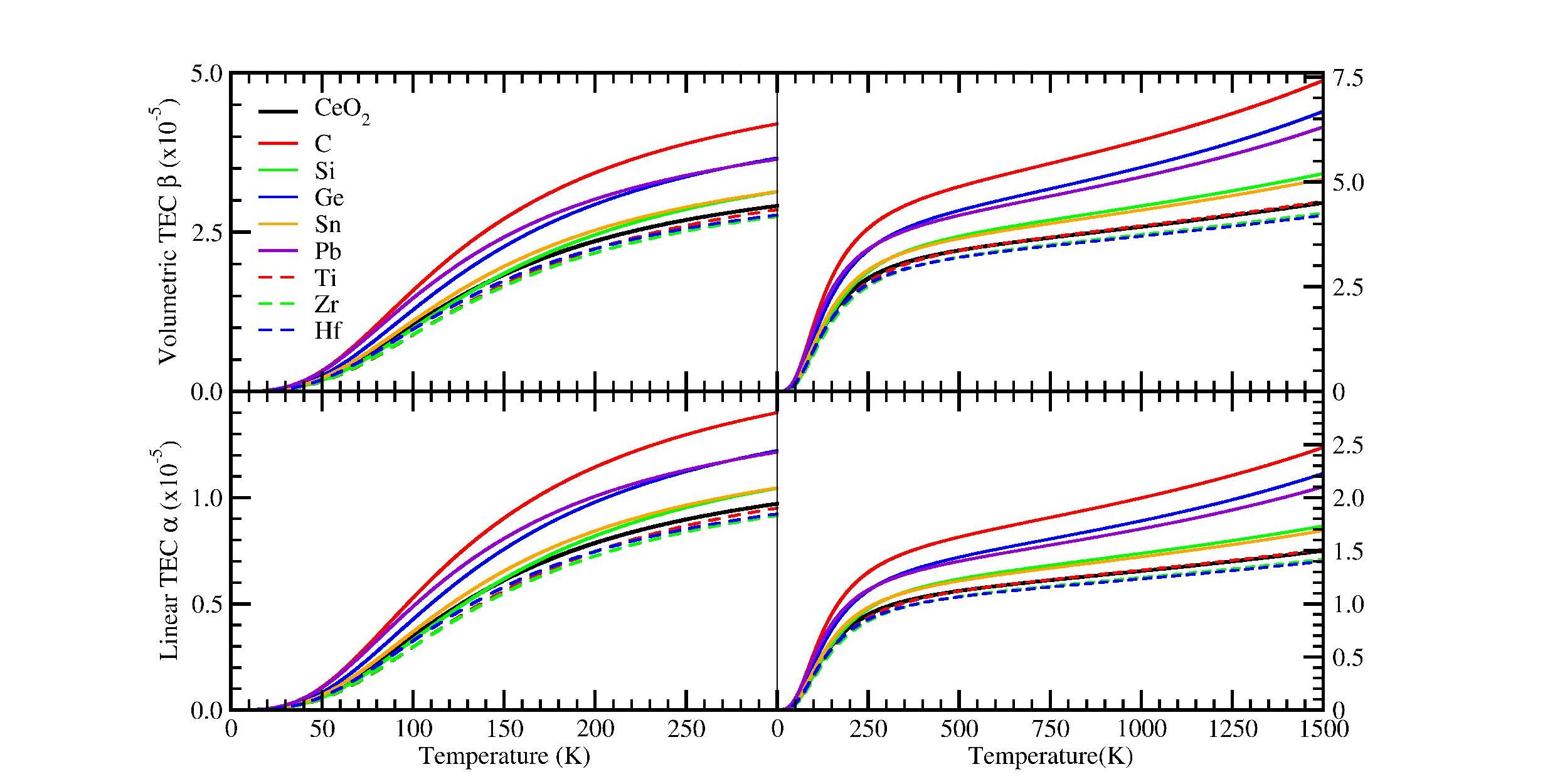}\\
\end{center}
  \caption[Thermal expansion coefficients for group IV doped \ce{CeO2}]{(Color online) Volumetric (top) and linear (bottom) TECs for LDA calculations. Dopant concentrations of $25$\% are used.}\label{fig:Subst_ThermEC_X4}
\end{figure}
\subsection[Thermal expansion coefficients and bulk moduli]{Thermal expansion coefficients and bulk moduli}\label{CeO2p4:ssc_ResultsBMTEC}
\indent We investigated the influence of group IV dopants on the mechanical properties of \ce{CeO2}, more specifically the thermal expansion coefficient (TEC) and the bulk modulus (BM). To reduce the computational cost, cells with a dopant concentration of $25$\% are used.\\
\indent The linear ($\alpha$) and volumetric ($\beta$) TECs of the doped systems can clearly be distinguished, as is seen in figure~\ref{fig:Subst_ThermEC_X4}. Whereas group IVa dopants result in a significant increase of the TEC, group IVb dopants result in a status quo or a very small decrease of the \ce{CeO2} TEC. In Table~\ref{table:Metal4subst_BM_TEC}, the calculated BM and linear TEC at $500$ K of doped \ce{CeO2} are compared, since both the TEC and the BM give a measure for how easily a material deforms under external conditions. Doping of \ce{CeO2} only has a small influence on the system's BM, except for C. As experimentally most often lower concentrations are used, even a smaller influence is expected.\\
\indent Similar as for the defect formation energies, an oscillation is seen for the BM and TEC values of group-IVa-doped ceria, showing the importance of filled $d$ or $f$ shells near the Fermi-level. For the group IVa elements, either the introduction of a filled $d$ or $f$ shell near the Fermi-level or the gap states in the O $2p$ -- Ce $4f$ band gap reduces the resistance to compression (\textit{i.e.} lower BM).\\
\indent Comparison of the BMs and TECs in Table~\ref{table:Metal4subst_BM_TEC} shows an inverse relation between the two: an increase in the BM corresponds to a decrease in the TEC, and vice versa. However, the relative positions of Si and Sn appear reversed, as do those of Ge and Pb. Examining the TEC curves presented in Fig.~\ref{fig:Subst_ThermEC_X4} shows that around $250$ K the order of the Si and Sn curves switches (the same for Ge and Pb), restoring the inverse behavior of the TEC and BM for the group IVa elements at lower temperatures. Because the BM is calculated at $0$ K the trend of the inverse behavior is found to be a general one. These findings agree with our expectations: First, since the BM is a measure of the resistance of a material against uniform compression/expansion, it stands to reason that a large BM will lead to a small TEC. Second, it is well known that both the BM and cohesive energy relate to the melting temperature, implying that the BM and cohesive energy correlate as well.\cite{Kittel_SolidStatePhysBook} On the other hand, the linear TEC $\alpha$ and the cohesive energy show an inverse correlation.\cite{TsuruY:JCeramJpn2010} As a result, an inverse correlation between the BM and linear TEC $\alpha$ is expected.\\
\indent For practical applications, \textit{e.g.} to reduce interfacial strain in a layered system through BM matching of the different layers,\cite{VandeVeldeNigel:EurJInorChem2010} a linear interpolation of the presented BM could be used to obtain a first order approximation of the optimal dopant concentration.\cite{VanpouckeDannyEP:2012aApplSurfSci} For the elements of group IVb, Table~\ref{table:Metal4subst_BM_TEC} shows a BM that is only slightly larger than that of \ce{CeO2} and that increases with the dopant atomic number. The group IVa elements show a more complex behavior (\textit{cf.} above), and with the exception of Si, all lower the BM of \ce{CeO2}. As a result, group IV elements are not suitable as dopants if an increase of the \ce{CeO2} BM is required. However, group IVa elements could be beneficial to obtain a lowered BM. In addition, the nature of the group IVa elements (\textit{cf.} Sec.~\ref{CeO2p4:ssc_ResultsEf}) may limit the effects to the grain boundaries or interfaces, allowing one to create a BM gradient using group IVa doped \ce{CeO2}.

\begin{figure}
\begin{center}
  \includegraphics[width=8cm,keepaspectratio=true]{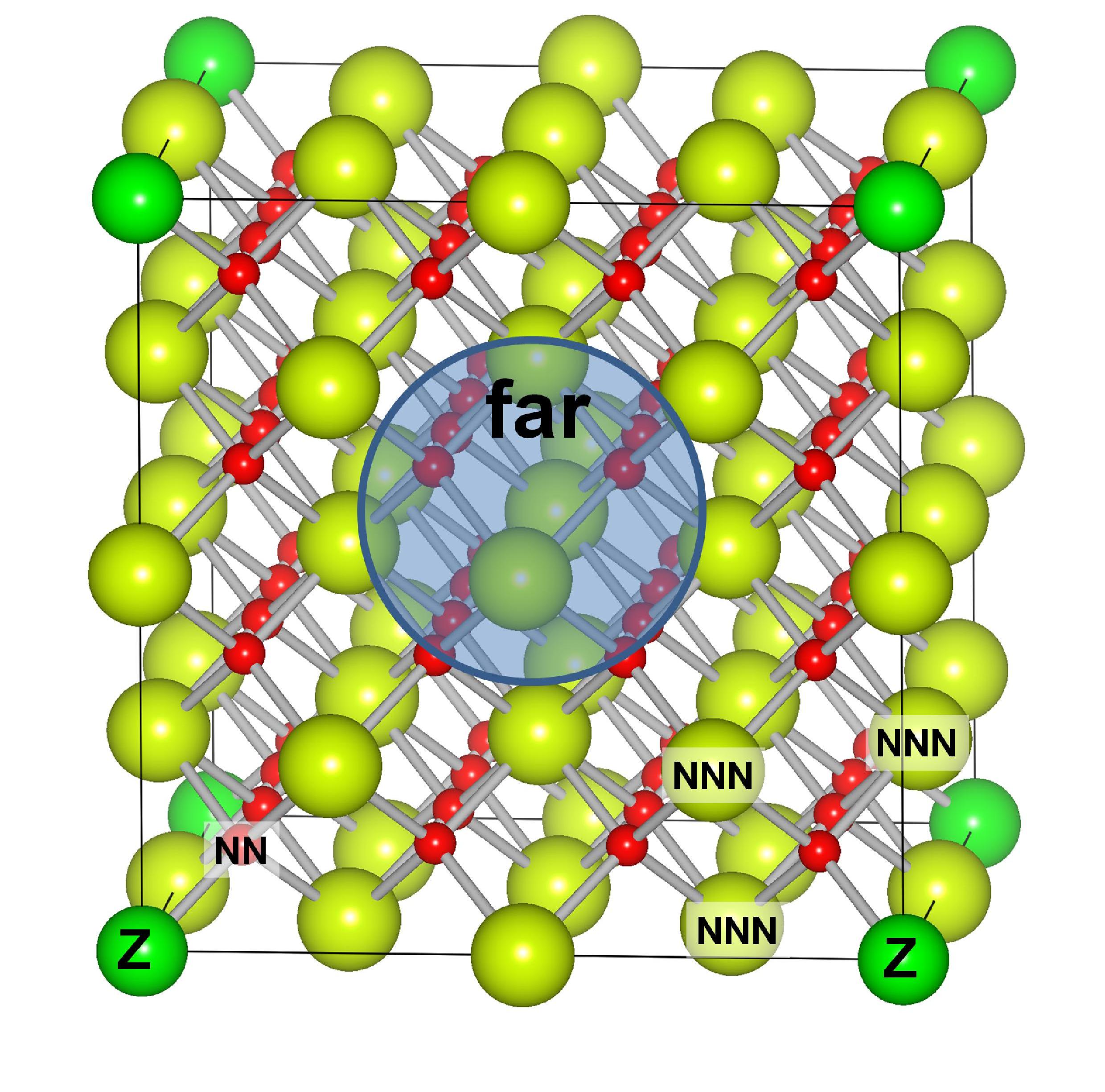}\\
\end{center}
  \caption[NN and NNN definition in a c222 supercell]{(Color online) Ball-and-stick model of a c222 \ce{CeO2} supercell doped with $3.125$\% of a group IV element. The red/yellow/green balls give the positions of the O/Ce/dopant(M) atoms. The nearest neighbor O atom in one octant is labeled NN, while the next nearest neighboring Ce atoms in a single octant are labeled NNN. Atoms `\emph{far}' away from the dopant are indicated in the blue disc.}\label{fig:NNdef}
\end{figure}
\begin{table}[!tbp]
\caption[Hirshfeld-I charges in \ce{Ce_{1-$x$}Z_{$x$}O2}, with Z a group IV element]{Hirshfeld-I charges in \ce{Ce_{1-$x$}Z_{$x$}O2}, with Z a group IV element.$^a$}\label{table:Metal4subst_HIRIcharge}
\begin{ruledtabular}
\begin{tabular}{l|ccccc}
& $M$ & \multicolumn{2}{c}{O} &  \multicolumn{2}{c}{Ce} \\
&   & NN & far & NNN & far \\
\hline
\ce{CeO2} & $-$ & \multicolumn{2}{c}{$-1.41$} & \multicolumn{2}{c}{$2.81$}\\
\hline
C  & $0.94$ & $-1.17$ & $-1.41$ & $2.79$ & $2.81$ \\
Si & $2.23$ & $-1.30$ & $-1.41$ & $2.79$ & $2.81$ \\
Ge & $2.33$ & $-1.31$ & $-1.410$ & $2.79$ & $2.81$ \\
Sn & $2.64$ & $-1.37$ & $-1.41$ & $2.80$ & $2.81$ \\
Pb & $2.49$ & $-1.36$ & $-1.41$ & $2.80$ & $2.81$ \\
\hline
Ti & $2.65$ & $-1.37$ & $-1.41$ & $2.80$ & $2.81$ \\
Zr & $2.97$ & $-1.42$ & $-1.41$ & $2.81$ & $2.81$ \\
Hf & $3.03$ & $-1.42$ & $-1.39$ & $2.80$ & $2.80$
\end{tabular}
\end{ruledtabular}
\begin{flushleft}
$^{a}$ Calculated Hirshfeld-I atomic charges for \ce{Ce_{1-$x$}Z_{$x$}O2} with $x=3.125$\% using LDA charge density distributions. All values are in $e$. The first column gives the atomic charge of the dopant $Z$. With regard to the dopant atom the shell of nearest and next nearest neighboring atoms (NN and NNN) consist entirely of O and Ce, respectively. The respective positions are shown in Fig.~\ref{fig:NNdef} The atomic charges of these atoms are compared to the charges of O and Ce atoms at a longer distance (indicated as `far').\\
\end{flushleft}
\end{table}
\subsection{Atomic charges and charge transfer}\label{CeO2p4:ssc_ResultsHI}
\indent The introduction of dopants in \ce{CeO2} not only influences the electronic structure at the level of bands and the density of states, but also at the level of the local charge density distribution. To investigate this effect we calculate the atomic charges in the doped systems using the Hirshfeld-I method.\cite{BultinckP:2007JCP_HirRef, VanpouckeDannyEP:2012bJComputChem, VanpouckeDannyEP:2012cJComputChem}
Table~\ref{table:Metal4subst_HIRIcharge} shows the calculated charges on the dopant elements. The charge of the O and Ce atoms in pure \ce{CeO2} is shown as reference.\\
\indent The atomic charges for all dopants show an increase of no more than $0.05 e$ when increasing the dopant concentration from $3.125$\% to $25$\%. This is a first indication that the influence of the dopants on the electron density distribution is quite localized. Mainly the atomic charge of the nearest neighbor O atoms changes. The atomic charges of the next nearest neighboring Ce atoms change very little due to doping (\textit{cf.} Fig.~\ref{fig:NNdef} and Table~\ref{table:Metal4subst_HIRIcharge}). The O atoms farther  from the dopant in the c222 cell show atomic charges that differ only slightly from those in pure \ce{CeO2}.\\
\indent Although all dopants are tetravalent the atomic charges vary significantly. The difference in atomic charge of the dopants and Ce is almost entirely compensated by the change in atomic charge on the nearest neighbor O atoms, showing that even the nearby O-Ce bonds are barely affected by these dopants. For all group IV elements the amount of non-compensated charge is less than $0.05 e$ per O atom in the nearest neighbor shell.\\
\indent All dopants behave similarly, with carbon being the sole exception. This may be an indication of the much weaker bonds of the C atom with the surrounding O atoms than the other dopants, which is in agreement with the fact that the small C atom prefers short bonds. The small size of \ce{C} (Shannon crystal radius\cite{Shannon:ACSA1976,Shannon:table} of $0.29$($0.3$)\AA\ for four (six) coordinated \ce{C}) might make it very suited as an interstitial dopant or as substitutional dopant on an oxygen site, which are interesting scenarios for a study focussing on \ce{C} doping of \ce{CeO2}, but are beyond the scope of this work.\\
\indent Note, however, that even in the C case the effects remain localized to the C dopant and the surrounding O atoms in the nearest neighbor shell. The fact that the C atom loses roughly two electrons less (\textit{i.e.} the two electrons remain localized on the C atom) than the other dopants and Ce to the surrounding O atoms is in line with the recent findings of Hellman \textit{et al.}\cite{HellmanO:PhysRevLett2012} They observed that for reduced C-doped ceria the two electrons, which provided the formal charge of -II on the removed O atom, localize in C $p$ orbitals.\\
\indent From the results in Table~\ref{table:Metal4subst_HIRIcharge}, it appears that the group IV dopants cause no significant changes in the atomic charges beyond the nearest neighbor shell.

\section{Conclusion}
\indent Using \textit{ab initio} DFT calculations we investigated the influence of tetravalent doping on the properties of fluorite \ce{CeO2}. Based on the calculated defect formation energies we conclude that Ce-substitution with group IVa elements is rather unlikely, while group IVb elements are good substituents for tuning the \ce{CeO2} fluorite lattice. This qualitative result is shown to be independent of the used functional and independent of the inclusion of a Coulomb correction.\\
\indent Investigation of the DOS and LDOS for both high and low dopant concentrations shows that the presence of states in the O $2p$--Ce $4f$ band gap are detrimental for the system. Low dopant concentration calculations show that group IVa dopants introduce localized gap states due to the overlap of their valence $s$ states with the O $2p$ states of the surrounding O atoms. These gap states are connected to occupied atomic states below the O $2p$ band. These gap states also provide the means for excess electrons in reduced ceria to localize on the dopant. In case of the IVb elements the gap state is always located above the unoccupied Ce $4f$ band resulting in a much reduced influence on the \ce{CeO2} properties.\\
\indent Comparison of the bulk modulus (BM) and the thermal expansion coefficient (TEC) shows that an inverse relation between these properties exists. Also here, group IVa and IVb elements differentiate, with group IVb elements resulting only in a slight increase of the BM, while a decrease in the BM is observed for group IVa dopants.\\
\indent The influence of group IV dopants on the charge distribution of the systems is very limited, and it is shown that, despite significant changes at the nearest neighbor atoms, no significant changes are observed farther away.\\
\indent Throughout all results presented, the impact of the filled $d$ and $f$ shells located near the Fermi level is observed as oscillations in the presented trends that coincide with the introduction of a new shell. This shows that not only the outer valence electrons play a role, but also filled shells can promote significant modifications of the expected results.\\

\begin{appendices}
\section{Computational details}
\label{CeO2p4:sc_theormeth}
\indent We perform \textit{ab-initio} spin-polarized DFT calculations using the PAW method as implemented in the Vienna \textit{ab-initio} Package (\textsc{vasp}) program. The LDA functional as parameterized by Ceperley and Alder and the GGA functional as constructed by Perdew, Burke and Ernzerhof (PBE) are used to model the exchange and correlation behavior of the electrons.\cite{Blochl:prb94, Kresse:prb99, CA:prl1980, PBE_1996prl, Kresse:prb93, Kresse:prb96} This is sufficient for the mechanical and structural properties studied. To asses the influence of the strong on-site Coulomb repulsion between the Ce $4f$ electrons on electronic properties such as defect formation energy and density of states, calculations are also performed within the DFT+U level framework, as formulated by Duradev et al.\cite{DuradevSL:PhysRevB1998} In different DFT+U studies, focussed on finding an optimum value for the Hubbard U parameter for ceria, a value for U of $5$--$6$ and $4$--$5$ eV is suggested for LDA and GGA, respectively.\cite{Loschen:prb2007, AnderssonDA:2007bPhysRevB, castleton:jcp2007, DaSilva:prb2007b} For this reason we have chosen to use an on-site Coulomb repulsion with U$=5$ eV for the Ce $4f$ electrons, for both LDA+U and PBE+U calculations. The plane wave kinetic energy cutoff is set to $500$ eV.\\
\indent Symmetric cells, containing a single substituent per cell are used to simulate a homogeneous distribution of dopants. Note that to have a clearer picture of the observed trends, C is also substituted at a Ce site and not at an O site, although this would be more in line with experimental expectations, as carbides of several rare earth elements, including cerium, are known.\cite{SpeddingFH:JAmChemSoc1958, AdachiGinYa:JLCM1973, SakaiTetsuo:JLCM1981} For all systems, the fluorite symmetry (space group $Fm\bar{3}m$) is maintained allowing for straightforward correlation of the observed trends and the dopant atomic properties. The cells used are the fluorite cubic $1\times 1\times 1$ cell with $12$ atoms (c$111$), the primitive $2\times 2\times 2$ cell with $24$ atoms (p$222$), the primitive $3\times 3\times 3$ cell with $81$ atoms (p$333$) and the cubic $2\times 2\times2$ cell with $96$ atoms (c$222$). Replacing a single Ce atom with a group IV element in these specific \ce{CeO2} supercells results in dopant concentrations of $25, 12.5, 3.7037,$ and $3.125$ \%, respectively.\\
\indent The valence electron configurations considered in these systems are $5s^2 5p^6 6s^2 4f^1 5d^1$ for Ce and $2s^2 2p^4$ for O. For the group IVa dopants C and Si only the $4$ $s$ and $p$ valence electrons are considered, while $10$ additional $d$ electrons are included for Ge, Sn and Pb. For the group IVb elements, the valence electron configurations considered are $3p^6 4s^2 3d^2$, $4s^2 4p^6 5s^2 4d^2$, and $5p^6 6s^2 5d^2$ for Ti, Zr, and Hf, respectively. The crystal structure information of the reference bulk materials used for calculating defect formation energies is presented in Table~\ref{table:Suppl1_GeomBulkMetal}, since many of these systems can have more than one (meta-)stable crystal structure.\\
\indent Monkhorst-Pack special $k$-point grids are used to sample the Brillouin zone.\cite{Monkhorst:prb76} For the two smaller cells we use an $8\times 8\times8$ $k$-point grid while for the two large cells a $4\times 4\times4$ $k$-point grid is used.\\
\indent To optimize the structures, a conjugate gradient method is used. During relaxation both atomic positions and cell-geometry are allowed to change simultaneously. Convergence is obtained when the difference in energy between subsequent steps is smaller than $1.0\times10^{-6}$ eV.\\
\indent The thermal expansion coefficients are calculated as the numerical derivative of V(T) data. In turn, these V(T) data are obtained through minimization of the thermal non-equilibrium Gibbs function, which is calculated using the quasi-harmonic Debye approximation, and is implemented as a module in our HIVE code.\cite{Maradudin:TheoryLattDynHarmApprox1971, BlancoMA:JMolStruc1996, FranciscoE:PRB2001, HIVE_REFERENCE}
The bulk modulus is calculated by fitting $E(V)$ data from fixed volume calculations to the third order isothermal Birch-Murnaghan equation of state.\cite{MurnaghanFD:PNAS1944,BirchF:PhysRev1947}\\
\indent To calculate the atomic charges of the doped systems, we used the Hirshfeld-I approach.\cite{BultinckP:2007JCP_HirRef, VanpouckeDannyEP:2012bJComputChem, VanpouckeDannyEP:2012cJComputChem} Our implementation makes use of the grid stored pseudo-electron density distributions obtained from \textsc{vasp}. The atom centered spherical integrations are done using Lebedev-Laikov grids of $1202$ grid points per shell, and a logarithmic radial grid.\cite{BeckeAD:1987JCP, LebedevVI_grid:1999DokladyMath} The iterative scheme is considered converged when the largest difference in charge of a system atom is less than $1.0\times 10^{-5}$ electron in two consecutive iterations.
\begin{table}[!tbp]
\caption[Defect formation energy for group IV dopants with regard to ZO2 oxides]{Defect formation energy for group IV dopants at different concentrations, with respect to the dopant oxides mentioned.}\label{table:Metal4subst_energiesBIS}
\begin{ruledtabular}
\begin{tabular}{l|cccc}
 & \multicolumn{4}{c}{\Ef (eV)} \\
 &  25\% & 12.5\% & 3.704\% & 3.125\% \\
\hline
 & \multicolumn{4}{c}{LDA}  \\
\hline
CO$_2$ & $13.276$ & $13.495$ & $13.588$ & $13.589$   \\
SiO$_2$ & $6.343$ & $6.133$ & $6.105$ & $6.107$  \\
GeO$_2$ & $4.192$ & $3.922$ & $3.898$ & $3.893$  \\
SnO$_2$ & $1.760$ & $1.565$ & $1.550$ & $1.550$  \\
PbO$_2$ & $1.082$ & $0.876$ & $0.861$ & $0.864$  \\
\hline
TiO$_2$ & $2.019$ & $2.105$ & $2.100$ & $2.085$ \\
ZrO$_2$ & $0.419$ & $0.444$ & $0.437$ & $0.432$ \\
HfO$_2$ & $0.706$ & $0.740$ & $0.732$ & $0.730$ \\
\hline
 & \multicolumn{4}{c}{PBE} \\
\hline
CO$_2$ & $13.353$ & $13.556$ & $13.620$ & $13.633$  \\
SiO$_2$ & $6.727$ & $6.537$ & $6.520$ & $6.552$  \\
GeO$_2$ & $3.936$ & $3.698$ & $3.683$ & $3.695$  \\
SnO$_2$ & $1.714$ & $1.551$ & $1.546$ & $1.545$  \\
PbO$_2$ & $1.085$ & $0.913$ & $0.903$ & $0.905$  \\
\hline
TiO$_2$ & $2.164$ & $2.236$ & $2.234$ & $2.242$  \\
ZrO$_2$ & $0.554$ & $0.587$ & $0.588$ & $0.586$  \\
HfO$_2$ & $0.765$ & $0.798$ & $0.798$ & $0.798$  \\
\end{tabular}
\end{ruledtabular}
\end{table}

\section{Oxide based defect formation energies}
\label{CeO2p4:Appendix_sc_Ef}
\indent The defect formation energy of doped ceria can also be calculated with regard to the dopant ZO$_2$ oxide (note that some group IV dopants, such as Pb, have multiple oxide stoichiometries). We chose to use ZO$_2$ type oxides since this corresponds to the stoichiometry of \ce{CeO2} itself. Furthermore, since elements like Si have a large number of (meta-)stable oxide polymorphs with the ZO$_2$ stoichiometry, and since temperature and pressure effects are not included in the \textit{ab initio} calculations, we opt for a low pressure/temperature structure as found in Reference \onlinecite{MerrillLeo:JPhysChemRefData1982}. The oxide reference structures are shown in Table~\ref{table:Suppl1_GeomBulkMetal}.
\indent The defect formation energy, with regard to the oxides presented in Table~\ref{table:Suppl1_GeomBulkMetal}, is calculated as:
\begin{equation}
E_f^{ZO_2} = E_{Ce_{1-x}Z_xO_2} + (x-1)E_{CeO_2} - xE_{ZO_2},
\end{equation}
with $E_{Ce_{1-x}Z_xO_2}$ the total energy of the doped system, $E_{CeO_2}$ the total energy of bulk \ce{CeO2} and $E_{ZO_2}$ the bulk energy of the oxide ZO$_2$ ( with Z = C, Si, Ge, Sn, Pb, Ti, Zr, and Hf). The resulting energies are shown in Table~\ref{table:Metal4subst_energiesBIS}. The same general trends are observed as for the defect formation energy presented in the manuscript: a) little concentration dependence, b) large range of energies depending on the dopant element, c) none of the dopants improve the stability of \ce{CeO2}. The obtained energies, however, are lower due to the increased stability of the formed oxides. This means that for experiments, starting from the oxides, it will be easier to form group IV doped \ce{CeO2}, then if one would start with the bulk dopant material.

\end{appendices}

\section{Acknowledgement}
\indent This research was financially supported by FWO-Vlaanderen, project n$^{\circ}$ G. $0802.09$N. We also acknowledge the Research Board of Ghent University. All calculations were carried out using the Stevin Supercomputer Infrastructure at Ghent University.



\begin{thebibliography}{10}

\bibitem{TrovarelliA:CatalRev1996}
A.~Trovarelli, ``Catalytic properties of ceria and {CeO}$_2$-containing
  materials,'' {\em Catal. Rev.-Sci. Eng.},~38[4]~439--520 (1996).

\bibitem{KasparJ:CatalToday1999}
J.~Ka$\check{\mathrm{s}}$par, P.~Fornasiero, and M.~Graziani, ``Use of
  {CeO}$_2$-based oxides in the three-way catalysis,'' {\em Catal. Today},~50[2]~285--298 (1999).

\bibitem{FuQ:2003Science}
Q.~Fu, H.~Saltsburg, and M.~Flytzani-Stephanopoulos, ``{Active Nonmetallic Au
  and Pt Species on Ceria-Based Water-Gas Shift Catalysts},''
  {\em Science},~301[5635]~935--938 (2003).

\bibitem{SheYusheng:IJHE2009}
Y.~She, Q.~Zheng, L.~Li, Y.~Zhan, C.~Chen, Y.~Zheng, and X.~Lin, ``{R}are earth
  oxide modified {CuO/CeO}$_2$ catalysts for the water�gas shift reaction,''
  {\em Int. J. Hydrogen Energy},~34[21]~8929--8936 (2009).

\bibitem{DeLeitenburgC:JChemSocChemCommun1995}
C.~De~Leitenburg, A.~Trovarelli, F.~Zamar, S.~Maschio, G.~Dolcetti, and
  J.~Llorca, ``A novel and simple route to catalysts with a high oxygen storage
  capacity - the direct room-temperature synthesis of {CeO}$_2$-{ZrO}$_2$ solid
  solutions,'' {\em J. Chem. Soc.-Chem. Commun.},~21~2181--2182 (1995).

\bibitem{KundakovicLj:JCat1998}
L.~Kundakovic and M.~Flytzani-Stephanopoulos, ``{Cu- and Ag-Modified Cerium
  Oxide Catalysts for Methane Oxidation},'' {\em J. Catal.},~179[1]~203--221 (1998).

\bibitem{ManzoliMaela:CT2008}
M.~Manzoli, G.~Avgouropoulos, T.~Tabakova, J.~Papavasiliou, T.~Ioannides, and
  F.~Boccuzzi, ``{P}referential {CO} oxidation in {H}$_2$-rich gas mixtures over
  {Au}/doped ceria catalysts,'' {\em Catal. Today},~138[3-4]~239--243 (2008).

\bibitem{LiB:IJHE2010}
B.~Li, X.~Wei, and W.~Pan, ``Improved electrical conductivity of
  {Ce}$_{0.9}${Gd}$_{0.1}${O}$_{1.95}$ and {Ce}$_{0.9}${Sm}$_{0.1}${O}$_{1.95}$
  by co-doping,'' {\em Int. J. Hydrogen Energy},~35[7]~3018--3022 (2010).

\bibitem{VanpouckeDannyEP:2011PhysRevB_LCO}
D.~E.~P. Vanpoucke, P.~Bultinck, S.~Cottenier, V.~Van~Speybroeck, and
  I.~Van~Driessche, ``Density functional theory study of
  {La}$_{2}${Ce}$_{2}${O}$_{7}$: {D}isordered fluorite versus pyrochlore
  structure,'' {\em Phys. Rev. B},~84[5]~054110 9pp. (2011).

\bibitem{ParanthamanM:1997PhysC}
M.~Paranthaman, A.~Goyal, F.~List, E.~Specht, D.~Lee, P.~Martin, Q.~He,
  D.~Christen, D.~Norton, J.~Budai, and D.~Kroeger, ``{G}rowth of biaxially
  textured buffer layers on rolled-{Ni} substrates by electron beam
  evaporation,'' {\em Physica C},~275[3-4]~266--272 (1997).

\bibitem{OhSanghyun:PhysC1998}
S.~Oh, J.~Yoo, K.~Lee, J.~Kim, and D.~Youm, ``Comparative study on the crack
  formations in the {CeO$_2$} buffer layers for {YBCO} films on textured {Ni}
  tapes and {Pt} tapes,'' {\em Physica C},~308[1-2]~91--98 (1998).

\bibitem{PennemanG:2004EuroCeram}
G.~Penneman, I.~Van~Driessche, E.~Bruneel, and S.~Hoste, ``Deposition of
  {C}e{O$_2$} buffer layers and {YBa$_2$Cu$_3$O$_{7-\delta}$ superconducting
  layers using an aqueous sol-gel method},''pp.~501--504 in {Key
  Engineering Materials}, Vol.~264--268, {\em {Euro Ceramics VIII, Pts
  1-3.}} Edited by {Mandal, H and Ovecoglu, L}. {Turkish Ceram Soc; European Ceram
  Soc}, 2004.

\bibitem{TakahashiY:PhysC2004}
Y.~Takahashi, Y.~Aoki, T.~Hasegawa, T.~Maeda, T.~Honjo, Y.~Yamada, and
  Y.~Shiohara, ``Preparation of {YBCO} coated conductor on metallic tapes using
  an {MOD} process,'' {\em Physica C},~412-414[2]~905--909 (2004).

\bibitem{KnothKerstin:PhysC2005}
K.~Knoth, B.~Schlobach, R.~H{\"u}hne, L.~Schultz, and B.~Holzapfel,
  ``{La$_2$Zr$_2$O$_7$} and {Ce}-�{Gd}-�{O} buffer layers for {YBCO} coated
  conductors using chemical solution deposition,'' {\em Physica C},~426-431[2]~979--984 (2005).

\bibitem{VandeVeldeNigel:EurJInorChem2010}
N.~Van~de Velde, D.~Van~de Vyver, O.~Brunkahl, S.~Hoste, E.~Bruneel, and
  I.~Van~Driessche, ``{CeO$_2$ Buffer Layers for HTSC by an Aqueous Sol-Gel
  Method - Chemistry and Microstructure},'' {\em Eur. J. Inor. Chem.},2010[2]~233--241 (2010).

\bibitem{VyshnaviN:2012JMaterChem}
V.~Narayanan, P.~Lommens, K.~De~Buysser, D.~E.~P. Vanpoucke, R.~Huehne,
  L.~Molina, G.~Van~Tendeloo, P.~Van Der~Voort, and I.~Van~Driessche, ``Aqueous
  {CSD} approach for the growth of novel, lattice-tuned
  {L}a$_x${C}e$_{1-x}${O}$_{\delta}$ epitaxial layers,''
  {\em J. Mater. Chem.},~22[17]~8476--8483 (2012).

\bibitem{MogensenM:SolStateI2000}
M.~Mogensen, N.~M. Sammes, and G.~A. Tompsett, ``Physical, chemical and
  electrochemical properties of pure and doped ceria,'' {\em Solid State
  Ionics},~129[1-4]~63--94 (2000).

\bibitem{KudoT:JES1975}
T.~Kudo and H.~Obayashi, ``{Oxygen Ion Conduction of the Fluorite-Type
  Ce}$_{1-x}${L}n$_x${O}$_{2-x/2}${ (Ln = Lanthanoid Element)},'' {\em J.
  Electrochem. Soc.},~122[1]~142--147 (1975).

\bibitem{McBrideJR:JApplPhys1994}
J.~R. McBride, K.~C. Hass, B.~D. Poindexter, and W.~H. Weber, ``Raman and
  {X}-ray studies of {C}e$_{1 - x}${RE}$_x${O}$_{2 - y}$, where {RE=La, Pr, Nd,
  Eu, Gd, and Tb},'' {\em J. Appl. Phys.},76[4]~2435--2441 (1994).

\bibitem{KundakovicLj:ApplCatalA1998}
L.~Kundakovic and M.~Flytzani-Stephanopoulos, ``Reduction characteristics of
  copper oxide in cerium and zirconium oxide systems,''
  {\em Appl. Catal. A},171[1]~13--29 (1998).

\bibitem{RossignolS:JMaterChem1999}
S.~Rossignol, F.~Gerard, and D.~Duprez, ``Effect of the preparation method on
  the properties of zirconia-ceria materials,'' {\em J. Mater. Chem.}9[7]~1615--1620 (1999).

\bibitem{BeraP:ChemMater2002}
P.~Bera, K.~R. Priolkar, P.~R. Sarode, M.~S. Hegde, S.~Emura, R.~Kumashiro, and
  N.~P. Lalla, ``Structural {I}nvestigation of {C}ombustion {S}ynthesized
  {Cu/CeO}$_2$ {C}atalysts by {EXAFS} and {O}ther {P}hysical {T}echniques:
  {F}ormation of a {Ce}$_{1-x}${Cu}$_x${O}$_{2-\delta}$ {S}olid {S}olution,''
  {\em Chem. Mater.},~14[8]~3591--3601 (2002).

\bibitem{YamamuraH:JCSJ2003}
H.~Yamamura, H.~Nishino, K.~Kakinuma, and K.~Nomura, ``{Crystal phase and
  electrical conductivity in the pyrochlore-type composition systems,
  {Ln}$_2${Ce}$_2${O}$_7$ (Ln = La, Nd, Sm, Eu, Gd, Y and Yb)},'' {\em J. Ceram. Soc.
  Jpn},~111[12]~902--906 (2003).

\bibitem{WangX:JPhysChemB2005}
X.~Wang, J.~A. Rodriguez, J.~C. Hanson, D.~Gamarra, A.~Mart\'{i}nez-Arias, and
  M.~Fern\'{a}ndez-Garc\'{i}a, ``{Unusual Physical and Chemical Properties of
  Cu in Ce}$_{1-x}${Cu}$_x${O}$_2$ {O}xides,''
  {\em J. Phys. Chem. B},~109[42]~19595--19603 (2005).

\bibitem{FaggDP:JSolStateChem2006}
D.~Fagg, J.~Frade, V.~Kharton, and I.~Marozau, ``The defect chemistry of
  {Ce(Pr, Zr)O}$_{2-\delta}$,'' {\em J. Sol. State Chem.},~179[5]~1469--1477 (2006).

\bibitem{TiwariA:ApplPhysLett2006}
A.~Tiwari, V.~M. Bhosle, S.~Ramachandran, N.~Sudhakar, J.~Narayan, S.~Budak,
  and A.~Gupta, ``Ferromagnetism in {Co} doped {CeO}$_2$: {O}bservation of a
  giant magnetic moment with a high {C}urie temperature,'' {\em Appl. Phys.
  Lett.},~88[14]~142511 3pp. (2006).

\bibitem{NakamuraA:PureApplChem2007}
A.~Nakamura, N.~Masaki, H.~Otobe, Y.~Hinatsu, J.~Wang, and M.~Takeda,
  ``Defect-fluorite oxides {M}$_{1-y}${Ln}$_{y}${O}$_{2-y/2}$({Ln} =
  lanthanide; {M = Hf, Zr, Ce, U, Th}): {S}tructure, property, and
  applications,'' {\em Pure Appl. Chem.},~79[10]~1691--1729 (2007).

\bibitem{SongYQ:JApplPhys2007}
Y.~Q. Song, H.~W. Zhang, Q.~Y. Wen, H.~Zhu, and J.~Q. Xiao, ``Co doping effect
  on the magnetic properties of {CeO}$_2$ films on {Si}($111$) substrates,''
  {\em J. Appl. Phys.},~102[4]~043912 5pp. (2007).

\bibitem{VodungboB:ApplyPhysLett2007}
B.~Vodungbo, Y.~Zheng, F.~Vidal, D.~Demaille, V.~H. Etgens, and D.~H. Mosca,
  ``Room temperature ferromagnetism of {Co} doped {CeO}$_{2 - \delta}$ diluted
  magnetic oxide: {E}ffect of oxygen and anisotropy,''
  {\em Appl. Phys. Lett.},~90[6]~062510 3pp. (2007).

\bibitem{deBiasi:JAlloysCompd2008}
R.~de~Biasi and M.~Grillo, ``Evidence for clustering in {Cu}$^{2+}$-doped
  {CeO}$_2$,'' {\em J. Alloys Compd.},~462[1-2]~15--18 (2008).

\bibitem{SinghalRK:JPhysDApplPhys2011}
R.~K. Singhal, P.~Kumari, S.~Kumar, S.~N. Dolia, Y.~T. Xing, M.~Alzamora, U.~P.
  Deshpande, T.~Shripathi, and E.~Saitovitch, ``Room temperature ferromagnetism
  in pure and {Co}- and {Fe}-doped {CeO}$_2$ dilute magnetic oxide: effect of
  oxygen vacancies and cation valence,''
  {\em J. Phys. D: Appl. Phys.},~44[16]~165002 6pp. (2011).

\bibitem{ReddyBM:JPhysChemB2005}
B.~M. Reddy, A.~Khan, P.~Lakshmanan, M.~Aouine, S.~Loridant, and J.-C. Volta,
  ``Structural {C}haracterization of {N}anosized {CeO}$_2$-{SiO}$_2$,
  {CeO}$_2$-{TiO}$_2$, and {CeO}$_2$-{ZrO}$_2$ {C}atalysts by {XRD}, {R}aman,
  and {HREM} {T}echniques,'' {\em J. Phys. Chem. B},~109[8]~3355--3363 (2005).

\bibitem{ReddyBM:CatalSurvAsia2005}
B.~Reddy and A.~Khan, ``Nanosized {CeO}$_2$�{SiO}$_2$, {CeO}$_2$�{TiO}$_2$, and
  {CeO}$_2$�{ZrO}$_2$ {M}ixed {O}xides: {I}nfluence of {S}upporting {O}xide on
  {T}hermal {S}tability and {O}xygen {S}torage {P}roperties of {C}eria,'' {\em
  Catal. Surv. Asia},~9[3]~155--171 (2005).

\bibitem{RocchiniE:JCatal2000}
E.~Rocchini, A.~Trovarelli, J.~Llorca, G.~W. Graham, W.~H. Weber,
  M.~Maciejewski, and A.~Baiker, ``Relationships between
  {S}tructural/{M}orphological {M}odifications and {O}xygen {S}torage--{R}edox
  {B}ehavior of {S}ilica-{D}oped {C}eria,'' {\em J. Catal.},~194[2]~461--478 (2000).

\bibitem{RocchiniE:JCatal2002}
E.~Rocchini, M.~Vicario, J.~Llorca, C.~de~Leitenburg, G.~Dolcetti, and
  A.~Trovarelli, ``Reduction and {O}xygen {S}torage {B}ehavior of {N}oble
  {M}etals {S}upported on {S}ilica-{D}oped {C}eria,''
  {\em J. Catal.},~211[2]~407--421 (2002).

\bibitem{LinR:ApplCatal2003}
R.~Lin, M.-F. Luo, Y.-J. Zhong, Z.-L. Yan, G.-Y. Liu, and W.-P. Liu,
  ``Comparative study of {CuO/Ce}$_{0.7}${Sn}$_{0.3}${O}$_2$, {CuO/CeO}$_2$ and
  {CuO/SnO}$_2$ catalysts for low-temperature {CO} oxidation,'' {\em Appl.
  Catal., A},~255[2]~331--336 (2003).

\bibitem{AnderssonDA:2007bPhysRevB}
D.~A. Andersson, S.~I. Simak, N.~V. Skorodumova, I.~A. Abrikosov, and
  B.~Johansson, ``Theoretical study of {CeO}$_2$ doped with tetravalent ions,''
  {\em Phys. Rev. B},~76[17]~174119 10pp. (2007).

\bibitem{AnderssonDA:2007ApplPhysLett}
D.~A. Andersson, S.~I. Simak, N.~V. Skorodumova, I.~A. Abrikosov, and
  B.~Johansson, ``Redox properties of {CeO}$_2$--{MO}$_2$ {(M = Ti, Zr, Hf, or
  Th)} solid solutions from first principles calculations,'' {\em Appl. Phys.
  Lett.},~90[3]~031909 3pp. (2007).

\bibitem{TangYuanhao:PhysRevB2010}
Y.~Tang, H.~Zhang, L.~Cui, C.~Ouyang, S.~Shi, W.~Tang, H.~Li, J.-S. Lee, and
  L.~Chen, ``First-principles investigation on redox properties of $m$-doped
  \text{CeO}$_{2}$ $(m=\text{Mn},\text{Pr},\text{Sn},\text{Zr})$,'' {\em
  Phys. Rev. B},~82[12]~125104 9pp. (2010).

\bibitem{YashimaMasatomo:JPhysChemC2009}
M.~Yashima, ``{Crystal Structures of the Tetragonal Ceria--Zirconia Solid
  Solutions Ce$_x$Zr$_{1-x}$O$_2$ through First Principles Calculations ($0
  \leq x \leq 1$)},'' {\em J. Phys. Chem. C},~113[29]~12658--12662 (2009).

\bibitem{VanpouckeDannyEP:2012aApplSurfSci}
D.~E.~P. Vanpoucke, S.~Cottenier, V.~Van~Speybroeck, P.~Bultinck, and
  I.~Van~Driessche, ``Tuning of {CeO}$_2$ buffer layers for coated
  superconductors through doping,'' {\em Appl. Surf. Sci.},~260~32--35 (2012).

\bibitem{AnderssonDA:2007aPhysRevB}
D.~A. Andersson, S.~I. Simak, B.~Johansson, I.~A. Abrikosov, and N.~V.
  Skorodumova, ``Modeling of {CeO}$_2$ , {Ce}$_2${O}$_3$ , and {CeO}$_{2-x}$ in
  the {LDA+U} formalism,'' {\em Phys. Rev. B},~75[3]~035109 6pp. (2007).

\bibitem{GandugliaPirovanoMVeronica:SurfSciReports2007}
M.~V. Ganduglia-Pirovano, A.~Hofmann, and J.~Sauer, ``Oxygen vacancies in
  transition metal and rare earth oxides: {C}urrent state of understanding and
  remaining challenges,'' {\em Surf. Sci. Reports},~62[6]~219--270 (2007).

\bibitem{PanhansMA:SolStateIon1993}
M.~Panhans and R.~Blumenthal, ``A thermodynamic and electrical conductivity
  study of nonstoichiometric cerium dioxide,''
  {\em Solid State Ionics},~60[4]~279--298 (1993).

\bibitem{Kittel_SolidStatePhysBook}
C.~Kittel, {\em Introduction to Solid State Physics}; pp. 57. Edited by S. Johnson and P. McFadden.
\newblock John Wiley and Sons, Inc.: New York, 7~ed., 1996.

\bibitem{TsuruY:JCeramJpn2010}
Y.~Tsuru, Y.~Shinzato, Y.~Saito, M.~Shimazu, M.~Shiono, and M.~Morinaga,
  ``Estimation of linear thermal expansion coefficient from cohesive energy
  obtained by ab-initio calculation of metals and ceramics,'' {\em J. Ceram.
  Soc. Jpn.},~118[1375]~241--245 (2010).

\bibitem{BultinckP:2007JCP_HirRef}
P.~Bultinck, C.~{Van~A}lsenoy, P.~W. Ayers, and R.~Carb\'{o}-Dorca, ``Critical
  analysis and extension of the {H}irshfeld atoms in molecules,'' {\em J. Chem.
  Phys.},~126[14]~144111 9pp. (2007).

\bibitem{VanpouckeDannyEP:2012bJComputChem}
D.~E.~P. Vanpoucke, P.~Bultinck, and I.~Van~Driessche, ``{Extending Hirshfeld-I
  to bulk and periodic materials},'' {\em J. Comput. Chem.},~34[5]~405--417 (2013).

\bibitem{VanpouckeDannyEP:2012cJComputChem}
D.~E.~P. Vanpoucke, I.~Van~Driessche, and P.~Bultinck, ``{Reply to 'Comment on
  ``Extending Hirshfeld-I to bulk and periodic materials'' '},'' {\em J.
  Comput. Chem.},~34[5]~422--427 (2013).

\bibitem{Shannon:ACSA1976}
R.~D. Shannon, ``{Revised effective ionic radii and systematic studies of
  interatomic distances in halides and chalcogenides},'' {\em Acta Cryst.},~A32~751--767 (1976).

\bibitem{Shannon:table}
J.~D. Van~Horn, ``{Electronic Table of Shannon Ionic Radii},'' 2001.
\newblock downloaded from \url{http://v.web.umkc.edu/vanhornj/shannonradii.htm} 08/13/2010.

\bibitem{HellmanO:PhysRevLett2012}
O.~Hellman, N.~V. Skorodumova, and S.~I. Simak, ``{Charge Redistribution
  Mechanisms of Ceria Reduction},'' {\em Phys. Rev. Lett.},~108[13]~135504 4pp. 2012.

\bibitem{Blochl:prb94}
P.~E. Bl{\"o}chl, ``Projector augmented-wave method,''
{\em Phys. Rev. B},~50[24]~17953--17979 (1994).

\bibitem{Kresse:prb99}
G.~Kresse and D.~Joubert, ``From ultrasoft pseudopotentials to the projector
  augmented-wave method,'' {\em Phys. Rev. B},~59[3]~1758--1775 (1999).

\bibitem{CA:prl1980}
D.~M. Ceperley and B.~J. Alder, ``Ground state of the electron gas by a
  stochastic method,'' {\em Phys. Rev. Lett.},~45[7]~566--569 (1980).

\bibitem{PBE_1996prl}
J.~P. Perdew, K.~Burke, and M.~Ernzerhof, ``Generalized gradient approximation
  made simple,'' {\em Phys. Rev. Lett.},~77[18]~3865--3868 (1996).

\bibitem{Kresse:prb93}
G.~Kresse and J.~Hafner, ``Ab initio molecular dynamics for liquid metals,''
  {\em Phys. Rev. B},~47[1]~558--561 (1993).

\bibitem{Kresse:prb96}
G.~Kresse and J.~Furthm\"uller, ``Efficient iterative schemes for ab initio
  total-energy calculations using a plane-wave basis set,''
  {\em Phys. Rev. B},~54[16]~11169--11186 (1996).
  
\bibitem{DuradevSL:PhysRevB1998}
S.~L. Dudarev, G.~A. Botton, S.~Y. Savrasov, C.~J. Humphreys, and A.~P. Sutton,
  ``Electron-energy-loss spectra and the structural stability of nickel oxide:
  {A}n {LSDA+U} study,'' {\em Phys. Rev. B},~57[3]~1505--1509 (1998).
  
\bibitem{Loschen:prb2007}
C.~Loschen, J.~Carrasco, K.~M. Neyman, and F.~Illas, ``First-principles
  {LDA}$+${U} and {GGA}$+${U} study of cerium oxides: {D}ependence on the
  effective {U} parameter,'' {\em Phys. Rev. B},~75[3]~035115 8pp. (2007).

\bibitem{castleton:jcp2007}
C.~W.~M. Castleton, J.~Kullgren, and K.~Hermansson, ``Tuning {LDA}$+${U} for
  electron localization and structure at oxygen vacancies in ceria,'' {\em J.
  Chem. Phys.},~127[24]~244704 11pp. (2007).

\bibitem{DaSilva:prb2007b}
J.~L.~F. Da~Silva, ``Stability of the {Ce}$_{2}${O}$_{3}$ phases: {A DFT}$+${U}
  investigation,'' {\em Phys. Rev. B},~76[19]~193108 4pp. (2007).

\bibitem{SpeddingFH:JAmChemSoc1958}
F.~H. Spedding, K.~Gschneidner, and A.~H. Daane, ``{The Crystal Structures of
  Some of the Rare Earth Carbides},'' {\em J. Am. Chem. Soc.},~80[17]~4499--4503 (1958).

\bibitem{AdachiGinYa:JLCM1973}
G.-Y. Adachi, T.~Nishihata, and J.~Shiokawa, ``Rare-earth mixed dicarbides,''
  {\em J. Less-Common MET},~32[2]~301--306 (1973).

\bibitem{SakaiTetsuo:JLCM1981}
T.~Sakai, G.-Y. Adachi, T.~Yoshida, and J.~Shiokawa, ``Magnetic and electrical
  properties of rare earth dicarbides and their solid solutions,'' {\em J.
  Less-Common MET},~81[1]~91--102 (1981).

\bibitem{Monkhorst:prb76}
H.~J. Monkhorst and J.~D. Pack, ``Special points for brillouin-zone
  integrations,'' {\em Phys. Rev. B},~13[12]~5188--5192 (1976).
  
\bibitem{Maradudin:TheoryLattDynHarmApprox1971}
A.~A. Maradudin, E.~W. Montroll, G.~H. Weiss, and I.~P. Ipatova, {\em Theory of
  lattice dynamics in the harmonic approximation}.
\newblock New York: Academic press, 2~ed., 1971.

\bibitem{BlancoMA:JMolStruc1996}
M.~A. Blanco, A.~M. Pendas, E.~Francisco, J.~M. Recio, and R.~Franco,
  ``Thermodynamical properties of solids from microscopic theory:
  {A}pplications to {M}g{F}$_2$ and {A}l$_2${O}$_3$,'' {\em Theochem-J. Mol.
  Struct.},~368~245--255 (1996).

\bibitem{FranciscoE:PRB2001}
E.~Francisco, M.~A. Blanco, and G.~Sanjurjo, ``Atomistic simulation of
  {SrF}$_{2}$ polymorphs,'' {\em Phys. Rev. B},~63[9]~094107 9pp. (2001).

\bibitem{HIVE_REFERENCE}
D.~E.~P. Vanpoucke, ``{HIVE} v2.1.''
  \url{http://users.ugent.be/~devpouck/hive_refman/index.html} (2011).

\bibitem{MurnaghanFD:PNAS1944}
F.~D. Murnaghan, ``{T}he {C}ompressibility of {M}edia under {E}xtreme
  {P}ressures,'' {\em Proc. Natl. Acad. Sci. USA},~30[9]~244--247 (1944).

\bibitem{BirchF:PhysRev1947}
F.~Birch, ``{F}inite {E}lastic {S}train of {C}ubic {C}rystals,'' {\em Phys.
  Rev.},~71[11]~809--824 (1947).
  
\bibitem{BeckeAD:1987JCP}
A.~D. Becke, ``A multicenter numerical integration scheme for polyatomic
  molecules,'' {\em J. Chem. Phys.},~88[4]~2547--2553 (1988).

\bibitem{LebedevVI_grid:1999DokladyMath}
V.~I. Lebedev and D.~Laikov, ``{Q}uadrature formula for the sphere of $131$-th
  algebraic order of accuracy,'' {\em Doklady Mathematics},~59[3]~477--481 (1999).

\bibitem{MerrillLeo:JPhysChemRefData1982}
L.~Merrill, ``{Behavior of the AB$_2$-Type Compounds at High Pressures and High
  Temperatures},'' {\em J. Phys. Chem. Ref. Data},~11[4]~1005--1064 (1982).

\end{thebibliography}

\end{document}